# Superconductivity above 30 K achieved in dense scandium

Xin He [a,1,2,3], Changling Zhang[a,1,2], Zhiwen Li[1,2], Sijia Zhang[1], Shaomin Feng[1], Jianfa Zhao[1,2], Ke Lu[1,2] Baosen Bin[1,2], Yi Peng[1,2], Xiancheng Wang*[1,2], Jin Song[1], Luhong Wang[4], Saori I. Kawaguchi[5], Cheng Ji[6], Bing Li[6], Haozhe Liu[6], J.S. Tse[7], Changqing Jin*[1,2,3]

[1]*Beijing National Laboratory for Condensed Matter Physics, Institute of Physics, Chinese Academy of Sciences, Beijing 100190, China*
[2]*School of Physical Sciences, University of Chinese Academy of Sciences, Beijing 100190, China*
[3] *Songshan Lake Materials Laboratory, Dongguan 523808, China*
[4] *Shanghai Advanced Research in Physical Sciences, Shanghai 201203, China*
[5] *Japan Synchrotron Radiation Research Institute, SPring-8, Sayo-gun Hyogo 679-5198, Japan*
[6] *Center for High Pressure Science & Technology Advanced Research, Beijing 100094, China*
[7] *Department of Physics, University of Saskatchewan, Canada*

**Superconductivity is one of most intriguing quantum phenomena, and the quest for elemental superconductors with high critical temperature ($T_c$) is of great scientific significance due to their relatively simple material composition and the underlying mechanism. Here we report the experimental discovery of densely compressed scandium (Sc) becoming the first elemental superconductor with $T_c$ breaking into 30 K range, which is comparable to the $T_c$ values of the classic La-Ba-Cu-O or LaFeAsO superconductors. Our results show that $T_c^{onset}$ of Sc increases from ~3 K at around 43 GPa to ~32 K at about 283 GPa ($T_c^{zero}$ ~ 31 K), which is well above liquid neon temperature. Interestingly measured $T_c$ shows no sign of saturation up to the maximum pressure achieved in our experiments, indicating that $T_c$ might be even higher upon further compression.**

[a] These authors contributed equally to this paper
* Corresponding authors: wangxiancheng@iphy.ac.cn; jin@iphy.ac.cn

# Introduction

Searching for high-$T_c$ superconductors is one of most important research topics in physical sciences. Elemental superconductors attract special and growing attention due to the simplicity of their singular composition [1, 2]. About 20 elemental solids are known to show superconductivity at ambient pressure, of which niobium has the highest superconducting transition temperature $T_c \sim 9.2$ K [1] and its alloy NbTi has been widely used for its excellent superconducting performance [3]. High pressure is a powerful tool and has been playing an important role in exploring new elemental superconductors and tuning superconducting properties [2]. For simple elemental metals, high pressure usually suppresses $T_c$ because of the electronic band broadening and resulting reduction of the density of state (DOS) near the Fermi energy. Lattice stiffness is another important factor that $T_c$ goes down under pressure. However, these pressure effects on superconductivity may be altered by subtle structural and electronic changes associated with bonding or phase transitions. Pressure may also enable non-superconducting metals or even insulating solids at ambient pressure to host superconductivity at high pressure. For example, calcium is non-superconducting at ambient pressure, but exhibits a high $T_c$ of 25 K at 161 GPa[4], and insulating sulfur shows superconductivity with $T_c \sim 17$ K at 200 GPa[5]. All known elemental superconductors with $T_c$ near or above 20 K are realized under pressure, such as Li with $T_c \sim 16-20$ K at 43-48 GPa [6, 7] and yttrium with $T_c \sim 17$ K at 89 GPa [8]. Our recent study showed that elemental titanium reached $T_c$ of 26.2 K at pressure of 248 GPa [9].

It is indicated that both Ca and Ti show very high $T_c$ among elemental solids because of the pressure induced $s$-$d$ electron transition, which also drives a sequence of phase transitions[4, 9-14]. The structure instability associated with the pressure induced tendency and occurrence of phase transitions favors superconductivity because of the lattice softening enhanced electron-phonon coupling strength[9]. The crystal structure and physical properties of Sc metal under pressure was examined by previous studies [15-24], which showed that Sc has the *hcp* structure (Sc-I) at ambient conditions, and pressure generates four structural transitions at about 23, 104, 140, and 240 GPa, respectively, producing high-pressure phases of Sc II, Sc III, Sc IV, and Sc V, respectively [17-19]. Sc II phase crystallizes in an incommensurate composite structure comprising a body centered host structure and a *C* face centered guest structure [18], and Sc V phase has a hexagonal lattice consisting of 6 screw helical chains [17], while Sc III and Sc IV phases are not fully determined [19].

Previous work reported that Sc starts to show superconductivity at 21 GPa with $T_c$ ~ 0.35 K [21]. With further compression, $T_c$ reaches as high as 19.6 K at 107 GPa, which occurs just at the phase boundary of Sc II and Sc III. However, $T_c$ was reported to drop to ~10 K when pressure is further increased and Sc III phase appears [20]. Here, we report high pressure measurements up to 283 GPa to explore superconductivity in Sc. A maximal $T_c$ above 30 K was observed at the highest experimental pressure, which sets a new record among elemental superconductors. To date Sc is the only known elemental superconductor with $T_c$ breaking into 30 K temperature range.

**Experimental Section**

The electrical resistance measurements were performed by using the four probe Van der Pauw method for tiny specimen as described in the literatures [25, 26]. The pressure was calibrated *via* the shift of the first order Raman edge frequency from the diamond cutlet as shown in previous works [9, 27, 28]. The applied current is 100 μA. Diamond anvil cells were used to produce high pressures. A variant of anvils with double beveled culet size of 20/140/300 μm, 30/140/300 μm or 50/140/300 μm were adopted in the experiments. A plate of T301 stainless steel covered with mixture of *c*BN powder and epoxy as insulating layer was used as the gasket. A hole of approximately 15~30 μm in diameter depending on top culet size was drilled in the center of the gasket to serve as high pressure chamber. *h*BN powder was used as pressure transmitting medium that filled in the high pressure chamber. We used the ATHENA procedure to produce the specimen assembly[26]. Four Pt foils with thickness of approximately 0.5 μm as the inner electrode were deposited on the culet surface, after which cross shaped Sc specimens with side lengths ~10 μm × 10 μm and thickness of 1 μm were adhered on the electrodes and culet surface. Tens of specimens are prepared for the experiments. Diamond anvil cells were put into a MagLab system that provides synergetic extreme environments with temperatures from 300 to 1.5 K and magnetic fields up to 9 T for the transport measurements.

**Results and discussion**

Figure 1a shows the temperature dependence of electrical resistance at high pressure up to 215 GPa measured during the warming process for Sc Sample 1. The Sc metal starts to show superconductivity with $T_c$ above 3.3 K at 43 GPa. The $T_c$ monotonously rises to ~26.2 K at 215 GPa, which is much higher than the 8.2 K at 74 GPa reported by J. J. Hamlin *et. al.*[22] and 19.6 K at 107 GPa reported by Debessai *et. al.*[20]. Considering that $T_c$ shows an increasing tendency with further compression, we extended to higher pressures up to 283 GPa for Sample 2 and observed enhancement of $T_c$ in dense Sc metal. Temperature dependence of the electrical resistance for sample 2 under various pressures are shown in Figure 1b. With further increasing pressure, $T_c$ reaches 32 K at the maximum experimental pressure of 283 GPa as shown in Figure 2, where both cooling and warming curves are presented. It is evidenced in the experiments that cooling process usually gave rather higher transition temperature since it is not easy to reach thermal equilibrium with helium cooling rate. Alternatively the warming process can be subtly tuned to reach thermo equilibrium that approaches the intrinsic properties of the measured samples. We hence chose the Tc measured from warming process. For the warming process the onset critical temperature ($T_c^{onset}$) and the zero resistance critical temperature ($T_c^{zero}$) of superconducting transition at 283 GPa can be determined by the derivative of resistance with respect to temperature $dR/dT$ to be 32 K and 31 K, respectively, demonstrating a very narrow superconducting transition. The data for samples 3 and 4 shown in Figure S1(a,b) further confirm the superconductivity of Sc under high pressure with $T_c$ above 30 K.

Among elemental superconductors, niobium has the highest $T_c$ of 9.2 K at ambient pressure, while under high pressure several elements exhibit high $T_c$ near or above 20 K, such as Li ($T_c$ ~ 15-20 K at 48 GPa [6, 7]), Ca ($T_c$ ~ 25 K at 220 GPa [4]), Y ($T_c$ ~ 20 K near 100 GPa [8]), V ($T_c$ ~17 K at 120 GPa [29]), S ($T_c$ ~ 17 K at 220 GPa [5]) and Ti ($T_c$ ~ 26.2 K near 248 GPa [9]). Superconductivity above 30 K in elemental solids has never been reported before, and Sc is the first and so far only known elemental superconductor with $T_c$ breaking into the record-setting 30 K range, which is comparable to the $T_c$ of the classic LaBaCuO [30] and LaFeAsO superconductors [31].

To further probe superconductivity of dense Sc metal under high pressure, we measured transport properties under different magnetic fields for Sample 2. Figure 3a presents the electrical resistance measured at 283 GPa with applying magnetic fields. It is seen that the superconducting transitions are gradually suppressed by the magnetic field. Here, the temperature at 90% of normal state resistance was used to plot the data of $T_c^{90\%}$ versus magnetic field measured at 283 GPa as shown in Figure 3b, from which the upper critical field at zero temperature $\mu_0H_{c2}(0)$ can be estimated. A linear fitting of $H_{c2}(T)$ leads to the slope of $dH_{c2}/dT|_{Tc}$ of −2.65 T/K. Using the Werthamer-Helfand-Hohenberg (WHH) formula of $\mu_0H_{c2}(T) = -0.69 \times dH_{c2}/dT|_{Tc} \times T_c$, the $\mu_0H_{c2}(0)$ value controlled by orbital depairing mechanism in a dirty limit ($\mu_0H_{c2}^{Orb}(0)$) was calculated to be 58 T. The Ginzburg Landau (GL) formula of $\mu_0H_{c2}(T) = \mu_0H_{c2}(0) \times (1-(T/T_c)^2)$ is also used to estimate the upper critical field at zero temperature. As shown in Figure 3b, the fitting of $\mu_0H_{c2}(T)$ by GL formula gives a value of $H_{c2}(0) = 43.7$ T that is slightly smaller than $\mu_0H_{c2}^{Orb}(0)$. Using the obtained value of $H_{c2}(0)$, the GL coherence length was calculated to be $\xi = 27.4$ Å via $\mu_0H_{c2}(0) = \Phi_0/2\pi\xi^2$, where $\Phi_0 = 2.067 \times 10^{-15}$ Web is the magnetic flux quantum.

Combining our high pressure x ray experiments (Fig. S2) and the phase transition behaviors reported by previous works [17,18,19], we plot the pressure driven structure-superconductivity phase diagram shown in Figure 4. Five different crystal structures of Sc are identified when pressure varies from ambient pressure to 297 GPa. No superconductivity above 2 K was observed in Sc-I within the pressure range of 0-23 GPa. With increasing pressure, $T_c$ monotonously rises from 3.1 K at 43 GPa to 32 K at 283 GPa as shown from the data of thin foil samples from S1 to S4.

One remarkable interesting feature is the pressure dependence of $T_c$ shows non-saturating behavior, which suggests that $T_c$ could be further enhanced by further compression at higher pressure conditions. We notice that the monotonous rise of $T_c$ is different from the previous work reported by Debessai *et. al.* [20], where an abrupt $T_c$ drop is found at 111 GPa that is near the boundary between the Sc II and Sc III phases. To double check this discrepancy, one additional high pressure transport measurement was performed on powder Sc sample instead of thin film. As shown in Fig. S1(c,d), at 102 GPa there exist two superconducting transitions at 17.5 K and 14 K, respectively. When further increasing pressure, the low $T_c$ transition gradually shifts toward high temperature while the high $T_c$ transition disappears, which implies that there does exist a phase transition near 102 GPa. The low $T_c$ superconductivity should arise from the Sc III phase, which is consistent with the $T_c$ drop near the phase transition reported in the previous work [20]. Except for the data in the Sc II phase between 82~102 GPa, the pressure dependence of $T_c$ for the thin film and powder samples are well matched. Sc II phase adopts an incommensurate modulated host-guest structure with space group *I*4/*mcm*(γ) [18]. This incommensurate structure consisting of two interpenetrating sublattices, a body-centered host structure and a *C* face centered guest structure, along the crystallographic *c* axis. For powder Sc sample, the incommensurability increases with pressure [18] and $T_c$ increases accordingly [20]. It is speculated that the strain in the thin film samples should suppress the increase of incommensurability under high pressure relative to the powder sample and lead to the $T_c$ discrepancy. In the phases from Sc III to Sc V, $T_c$ rises consistently at rising pressure. Sc III and Sc IV phases are not fully resolved, although a recent theoretical study suggested that the *Ccca* 20 phase and *Cmca* 32 phase are likely candidates for the observed Sc III and Sc IV, respectively [19]. Sc V phase occurs when pressure exceeds 240 GPa, which adopts a hexagonal lattice (space group $P6_122$) consisting of 6 screw helical chains [17]. These results suggest that pressure induced structure instability plays a key role in driving rising $T_c$ in dense Sc, possibly driven by rising contribution of *d* orbital electronic states near the Fermi energy and the lattice softening associated with the phase transitions, both of which strengthen the electron phonon coupling.

We also have performed electron phonon coupling density functional calculations to confirm the experimentally observed superconductivity as seen in Fig. S3. Details of the calculations are given in supplementary materials. The predicted $T_c$ at 240 GPa varies from 30 to 33 K with empirical Coulomb repulsion parameter from 0.13 to 0.1. The results are similar to those reported in ref. 32. A valence electronic configuration Sc of $3s^{1.56}3p^{5.25}3d^{2.42}4s^{0.25}$ was obtained from Mulliken population analysis of the crystal orbitals calculated using atom-centered localized basis set [33]. It is obvious the substantial charge migration from the Sc 4*s* to the 3*d* orbitals is responsible for the strong electron phonon coupling.

## Conclusions

In summary, we have investigated transport properties of Sc under high pressure up to 283 GPa, and we have found that $T_c$ is monotonously enhanced by pressure from 3.1 K at around 43 GPa to 32 K at around 283 GPa, following a sequence of high-pressure phase transitions from Sc II to Sc V. The increase of $T_c$ shows no sign of saturation up to the highest experimental pressure of 283 GPa, which indicates that $T_c$ of superconducting Sc still has room to go higher upon further compression.

Scandium is the only currently known elemental superconductors with the $T_c^{onset}$ breaking through 30 K at high pressure. The present results offer fresh insights for exploring high-$T_c$ superconductivity at ultrahigh pressures in diverse element solids.

During preparing the paper (arXiv:2303.01062) we became aware that an independent similar work (arXiv:2302.14378) has been carried out by J. J. Ying *et al*.[32].

**Acknowledgments:**

The works are supported by NSF& MOST of China through research Projects. JST wish to thank NSERC Canada for a Discovery Grant. We thank C.F.Chen for the discussions.

# References

[1]. Buzea C and Robbie K 2005 Superconductor Science & Technology 18 R1.

[2]. Hamlin J J 2015 Physica C-Superconductivity and Its Applications 514 59.

[3]. Scanlan R M, Malozemoff A P and Larbalestier D C 2004 Proceedings of the Ieee 92 1639.

[4]. Yabuuchi T, Matsuoka T, Nakamoto Y and Shimizu K 2006 Journal of the Physical Society of Japan 75 083703.

[5]. Gregoryanz E, Struzhkin V V, Hemley R J, Eremets M I, Mao H K and Timofeev Y A 2002 Physical Review B 65 064504.

[6]. Shimizu K, Ishikawa H, Takao D, Yagi T and Amaya K 2002 Nature 419 597.

[7]. Struzhkin V V, Eremets M I, Gan W, Mao H K and Hemley R J 2002 Science 298 1213.

[8]. Hamlin J J, Tissen V G and Schilling J S 2007 Physica C-Superconductivity and Its Applications 451 82.

[9]. Zhang C L, He X, Liu C, Li Z W, Lu K, Zhang S J, Feng S M, Wang X C, Peng Y, Long Y W, Yu R C, Wang L H, Prakapenka V, Chariton S, Li Q, Liu H Z, Chen C F and Jin C Q 2022 Nature Communications 13 5411.

[10]. Olijnyk H and Holzapfel W B 1984 Physics Letters A 100 191.

[11]. Akahama Y, Kawamura H and Le Bihan T 2002 Journal of Physics-Condensed Matter 14 10583.

[12]. Ahuja R, Dubrovinsky L, Dubrovinskaia N, Guillen J M O, Mattesini M, Johansson B and Le Bihan T 2004 Physical Review B 69 184102.

[13]. Oganov A R, Ma Y, Xu Y, Errea I, Bergara A and Lyakhov A O 2010 Proceedings of the National Academy of Sciences 107 7646.

[14]. Liu X, Jiang P, Wang Y, Li M, Li N, Zhang Q, Wang Y, Li Y-L and Yang W 2022 Physical Review B 105 224511.

[15]. Vohra Y K, Grosshans W and Holzapfel W B 1982 Physical Review B 25 6019.

[16]. Zhao Y C, Porsch F and Holzapfel W B 1996 Physical Review B 54 9715.

[17]. Akahama Y, Fujihisa H and Kawamura H 2005 Physical Review Letters 94 195503.

[18]. Fujihisa H, Akahama Y, Kawamura H, Gotoh Y, Yamawaki H, Sakashita M, Takeya S and Honda K 2005 Physical Review B 72 132103.

[19]. Zhu S C, Yan X Z, Fredericks S, Li Y-L and Zhu Q 2018 Physical Review B 98 214116.

[20]. Debessai M, Hamlin J J and Schilling J S 2008 Physical Review B 78 064519.

[21]. Wittig J, Probst C, Schmidt F A and Gschneidner K A 1979 Physical Review

Letters 42 469.

[22]. Hamlin J J and Schilling J S 2007 Physical Review B 76 012505.

[23]. Arapan S, Skorodumova N V and Ahuja R 2009 Physical Review Letters 102 085701.

[24]. Tsuppayakorn-aek P, Luo W, Pungtrakoon W, Chuenkingkeaw K, Kaewmaraya T, Ahuja R and Bovornratanaraks T 2018 Journal of Applied Physics 124 225901.

[25]. Li Z W, He X, Zhang C L, Wang X C, Zhang S J, Jia Y T, Feng S M, Lu K, Zhao J F, Zhang J, Min B S, Long Y W, Yu R C, Wang L H, Ye M Y, Zhang Z S, Prakapenka V, Chariton S, Ginsberg P A, Bass J, Yuan S H, Liu H Z and Jin C Q 2022 Nature Communications 13 2863.

[26]. Jia Y T, He X, Feng S M, Zhang S J, Zhang C L, Ren C W, Wang X C and Jin C Q 2020 Crystals 10 1116.

[27]. Zhang C L, He X, Li Z W, Zhang S J, Feng S M, Wang X C, Yu R C and Jin C Q 2022 Science Bulletin 67 907.

[28]. Zhang C L, He X, Li Z W, Zhang S J, Min B S, Zhang J, Lu K, Zhao J F, Shi L C, Peng Y, Wang X C, Feng S M, Yu R C, Wang L H, Prakapenka V B, Chariton S, Liu H Z and Jin C Q 2022 Materials Today Physics 27 100826.

[29]. Ishizuka M, Iketani M and Endo S 2000 Physical Review B 61 R3823.

[30]. Chu C W, Hor P H, Meng R L, Gao L, Huang Z J and Wang Y Q 1987 Physical Review Letters 58 405.

[31]. Kamihara Y, Watanabe T, Hirano M and Hosono H 2008 Journal of the American Chemical Society 130 3296.

[32]. Ying J J, Liu S Q, Lu Q, Wen X K, Gui Z G, Zhang Y Q, Wang X M, Sun J and Chen X H 2023 Physical Review Letters 130 256002.

[33]. Te Velde G and Baerends E J 1991 Physical Review B 44 7888.

**Figure Captions**

**Figure 1** Temperature dependence of the electrical resistance of elemental Sc metal measured at high pressures for (a) sample 1 and (b) sample 2.

**Figure 2** The resistance curves measured at 283 GPa in both cooling and warming processes, where the derivative of the resistance with respect to temperature d$R$/d$T$ for the warming process is plotted to clearly show the $T_c^{onset}$ and $T_c^{zero}$.

**Figure 3** (a) Temperature dependence of the electrical resistance of Sc metal measured at different magnetic fields under fixed pressure of 283 GPa. The dashed line marks the 90% of the normal state resistance. (b) Upper critical field $\mu_0 H_{c2}$ versus temperature at 283 GPa. The Ginzburg Landau fitting for the $H_{c2}$ (T) is shown by the solid lines. The star symbol represents the $H_{c2}$ (0) values calculated via the WHH model.

**Figure 4**: The phase diagram of superconducting transition temperature $T_c$ and crystal structure versus pressure for Sc. The measured results on all the 6 samples show consistent trends.

**Figure 1**

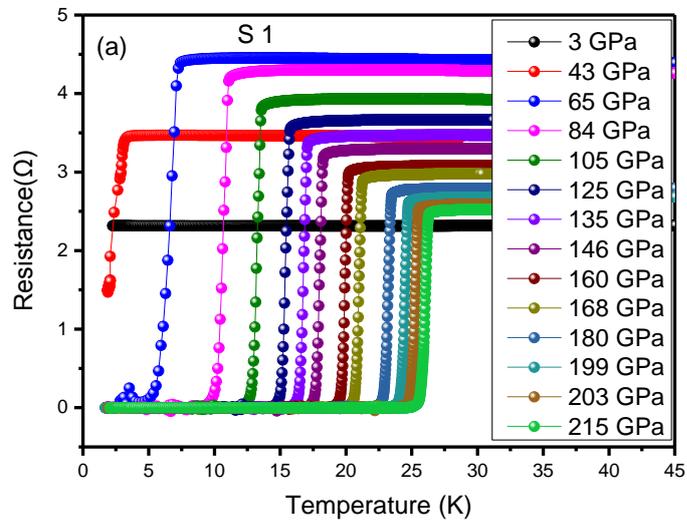

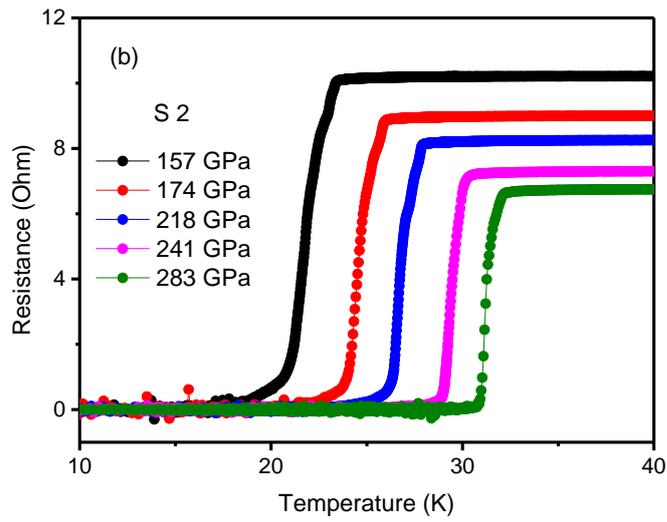

**Figure 2**

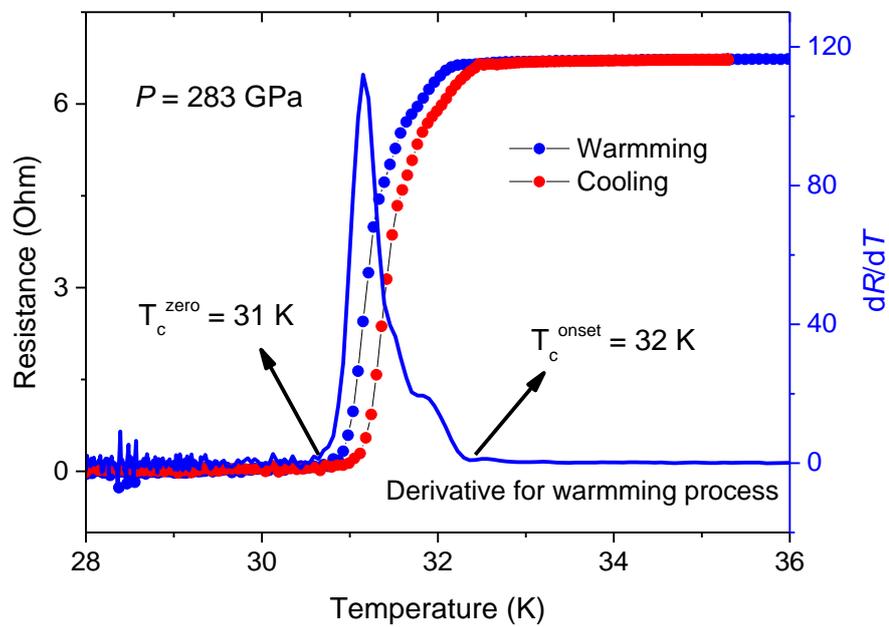

**Figure 3**

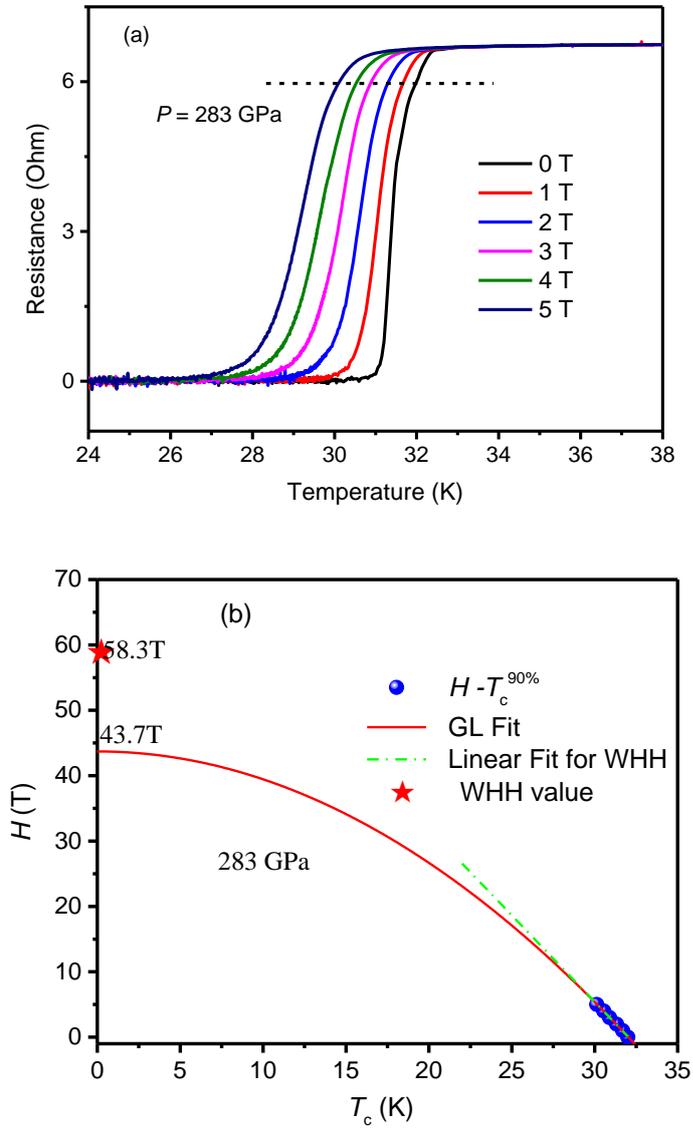

**Figure 4**

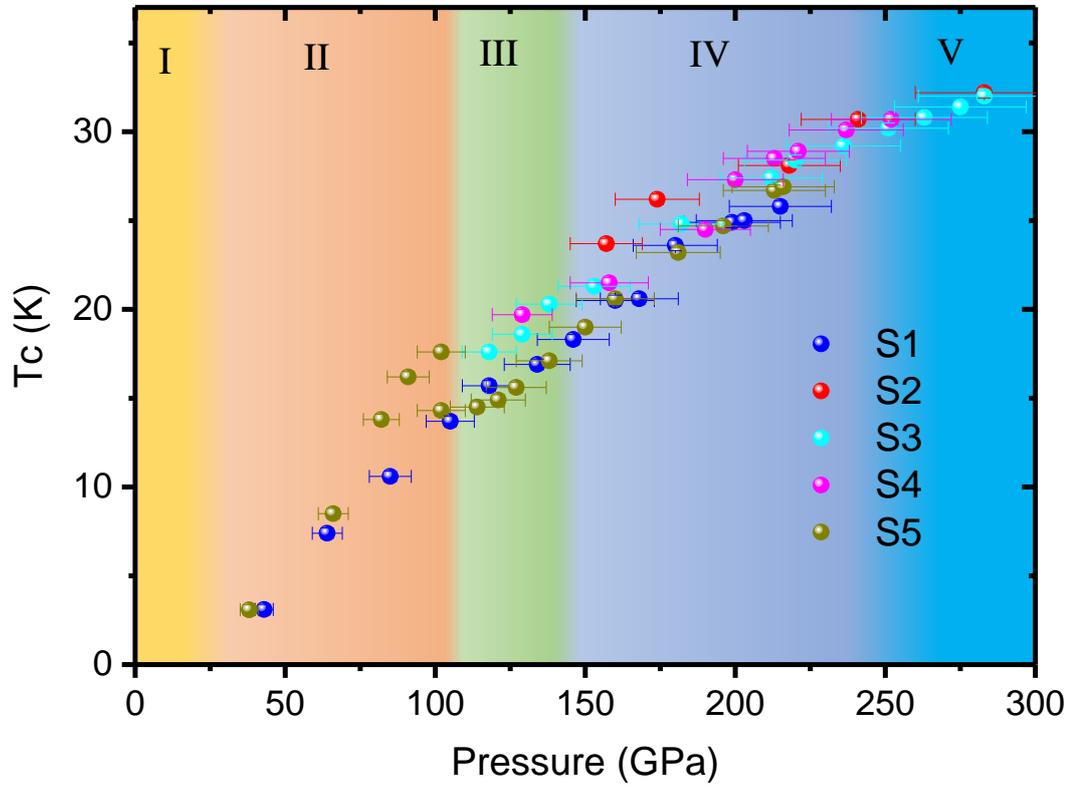

Supplementary Materials

# Superconductivity above 30 K achieved in dense scandium

Xin He [a,1,2,3], Changling Zhang [a,1,2], Zhiwen Li [1,2], Sijia Zhang [1], Shaomin Feng [1], Jianfa Zhao [1,2], Ke Lu [1,2] Baosen Bin [1,2], Yi Peng [1,2], Xiancheng Wang*[1,2], Jin Song [1], Luhong Wang [4], Saori I. Kawaguchi [5], Cheng Ji [6], Bing Li [6], Haozhe Liu [6], J.S. Tse [7], Changqing Jin [*,1,2,3]

[1] Beijing National Laboratory for Condensed Matter Physics, Institute of Physics, Chinese Academy of Sciences, Beijing 100190, China
[2] School of Physical Sciences, University of Chinese Academy of Sciences, Beijing 100190, China
[3] Songshan Lake Materials Laboratory, Dongguan 523808, China
[4] Shanghai Advanced Research in Physical Sciences, Shanghai 201203, China
[5] Japan Synchrotron Radiation Research Institute, SPring-8, Sayo-gun Hyogo 679-5198, Japan
[6] Center for High Pressure Science & Technology Advanced Research, Beijing 100094, China
[7] Department of Physics, University of Saskatchewan, Canada

## I. Experiments

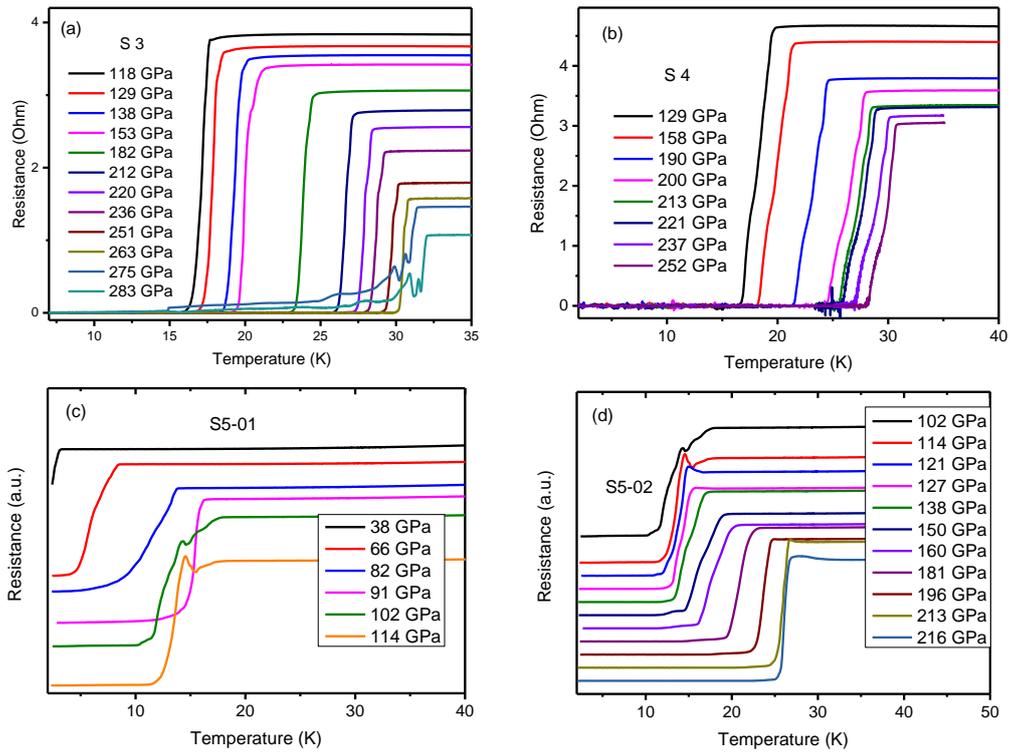

Fig. S1 (a-d) Temperature dependence of the electrical resistance of elemental Sc metal measured at high pressures for sample 3-5. Sample 3 and 4 are Sc films while sample 5 are Sc powder.

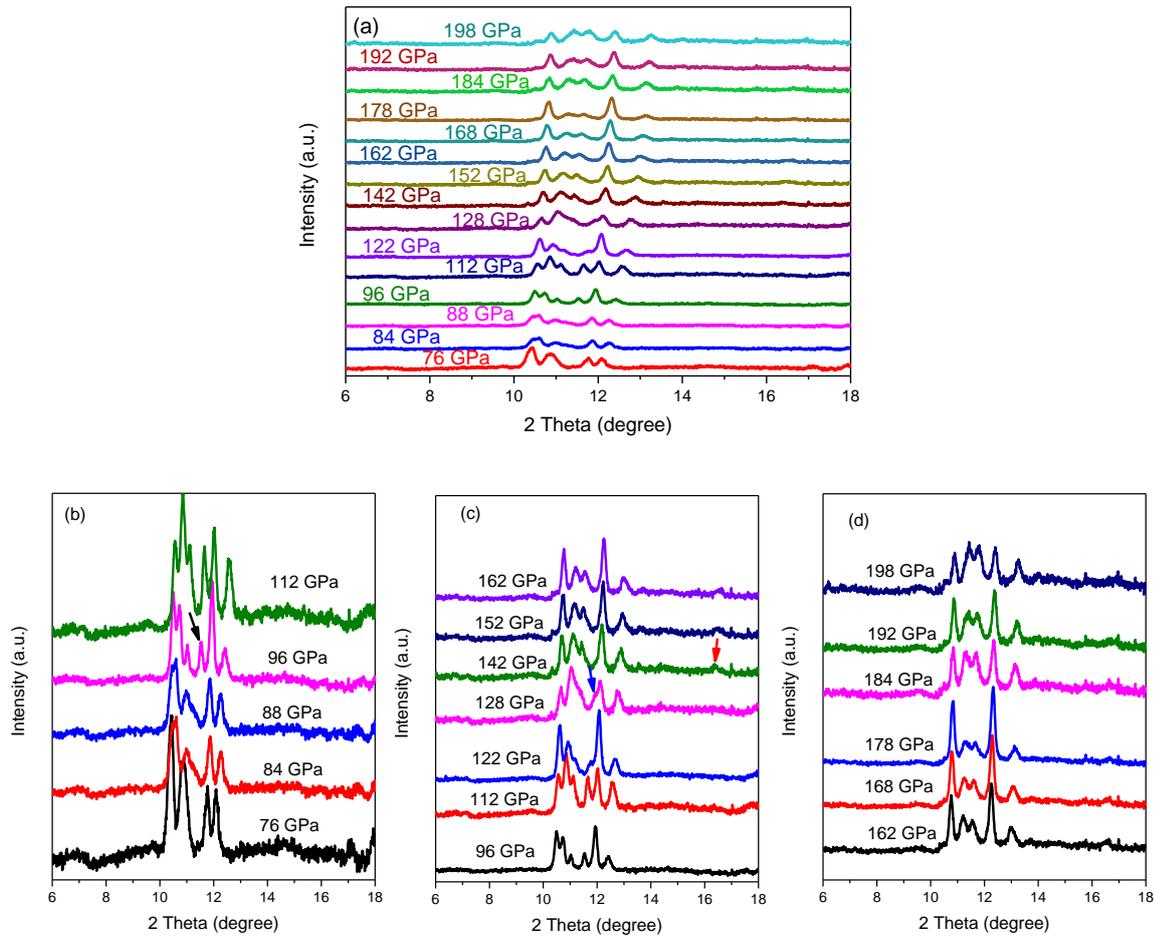

Fig. S2 (a-d) The *in-situ* high pressure synchrotron x-ray diffraction experiments were carried out at BL10XU, SPring-8, Japan, with the wavelength of 0.4126 Angstrom. As shown in Fig. S2(b), a new peak at around 2 Theta = 11.5 degree, marked by black arrow in the diffraction pattern collected at 96 GPa, suggests a phase transition from the Sc-II to Sc-III phase. Fig. S3(c) demonstrates a phase transition from Sc-III to Sc-IV occurs at about 142 GPa, which is suggested by the peak appearance at 2 Theta at about 16.4 degree and the peak disappearance at about 11.9 degree marked by red and blue arrows, respectively. In the pressure range from 142 to 198 GPa, the Sc-IV phase is stable as seen in Fig. S2(d).

## II. Theoretical calculations

The electronic structure, phonons and electron-phonon coupling on Sc-V at 240 GPa were calculated using the Quantum-Espresso (QE) package [S1] with the Optimized Norm-Conserving Vanderbilt (ONCV) PBE [S2] pseudopotential for Sc obtained from the QE website. The energy cutoff is 80 Ryd and the density cutoff is 640 Ryd. The *k*-point sets for SCF and charge density calculations were 12×12×6 and 24×24×12, respectively. Th *q*-point set for phonon calculations was 6×6×6. The computed band structure, phonon density of states (vDOS) and Eliashberg function ($\alpha^2 F(\omega)$) are shown in Fig. S3. The results are in very good agreement with those reported in ref. 32. It is interesting to note that the profiles of the vDOS and $\alpha^2 F(\omega)$ are very similar, as in the typical elemental superconductor Nb.

The valence orbital occupation of S-V at 240 GPa were calculation with the ADF suite [S3] using the PBE functional on the optimized structure described above. A triple-zeta augmented with polarization functions (TZ2P) basis set for Sc was used.

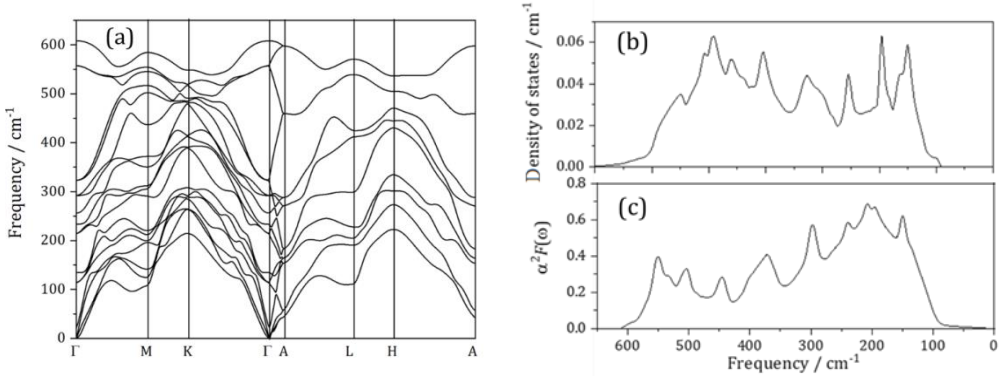

Fig. S3 (a-c) Calculated electronic band structure, phonon vibrational density of states and Eliashberg function for Sc-V at 240 GPa.

# References

[S1] P. Giannozzi, O. Andreussi, T. Brumme, O. Bunau, M. Buongiorno Nardelli, M. Calandra, R. Car, C. Cavazzoni, D. Ceresoli, M. Cococcioni, N. Colonna, I. Carnimeo, A. Dal Corso, S. de Gironcoli, P. Delugas, R. A. DiStasio Jr, A. Ferretti, A. Floris, G. Fratesi, G. Fugallo, R. Gebauer, U. Gerstmann, F. Giustino, T. Gorni, J Jia, M. Kawamura, H.-Y. Ko, A. Kokalj, E. Küçükbenli, M .Lazzeri, M. Marsili, N. Marzari, F. Mauri, N. L. Nguyen, H.-V. Nguyen, A. Otero-de-la-Roza, L. Paulatto, S. Poncé, D. Rocca, R. Sabatini, B. Santra, M. Schlipf, A. P. Seitsonen, A. Smogunov, I. Timrov, T. Thonhauser, P. Umari, N. Vast, X. Wu, S. Baroni, J.Phys.: Condens.Matter 29, 465901 (2017).

[S2] J. P. Perdew, K. Burke, and M. Ernzerhof, Phys. Rev. Lett. 77, 3865 (1996).

[S3] BAND 2023.1, SCM, Theoretical Chemistry, Vrije Universiteit, Amsterdam, The Netherlands, http://www.scm.com Optionally, you may add the following list of authors and contributors: P.H.T. Philipsen, G. te Velde, E.J. Baerends, J.A. Berger, P.L. de Boeij, M. Franchini, J.A. Groeneveld, E.S. Kadantsev, R. Klooster, F. Kootstra, M.C.W.M. Pols, P. Romaniello, M. Raupach, D.G. Skachkov, J.G. Snijders, C.J.O. Verzijl, J.A. Celis Gil, J. M. Thijssen, G. Wiesenekker, C. A. Peeples, G. Schreckenbach, T. Ziegler.